\documentclass[aps,twocolumn,10pt,groupedaddress]{revtex4-1}  
\usepackage{graphicx}  
\usepackage{subfigure}
\usepackage{dcolumn}   
\usepackage{bm}        
\usepackage{amssymb,amsmath}   

\begin{document}
\title{A sub-40 mHz linewidth laser based on a silicon single-crystal optical cavity}

\author{T. Kessler} 
\altaffiliation{member of QUEST Institute for Experimental Quantum Metrology, Bundesallee 100, 38116 Braunschweig, Germany}
\author{C. Hagemann}
\altaffiliation{member of QUEST Institute for Experimental Quantum Metrology, Bundesallee 100, 38116 Braunschweig, Germany}
\author{C. Grebing}
\author{T. Legero}
\author{U. Sterr}
\author{F. Riehle}
\affiliation{Physikalisch-Technische Bundesanstalt, Bundesallee 100, 38116 Braunschweig, Germany}
\author{M.J. Martin}
\author{L. Chen}
\altaffiliation[current address: ]{Wuhan Institute of Physics and Mathematics,
 Chinese Academy of Sciences, Wuhan, 430071, China}
\author{J. Ye}
\affiliation{JILA, National Institute of Standards and Technology and University of Colorado, Department of Physics, 440 UCB, Boulder, Colorado 80309, USA}

\begin{abstract}
State-of-the-art optical oscillators based on lasers frequency stabilized to high finesse optical cavities are limited by thermal noise that causes fluctuations of the cavity length. Thermal noise represents a fundamental limit to the stability of an optical interferometer and plays a key role in modern optical metrology. We demonstrate a novel design to reduce the thermal noise limit for optical cavities by an order of magnitude and present an experimental realization of this new cavity system, demonstrating the most stable oscillator of any kind to date. The cavity spacer and the mirror substrates are both constructed from single crystal silicon and operated at 124~K where the silicon thermal expansion coefficient is zero and the silicon mechanical loss is small. The cavity is supported in a vibration-insensitive configuration, which, together with the superior stiffness of silicon crystal, reduces the vibration related noise. With rigorous analysis of heterodyne beat signals among three independent stable lasers, the silicon system demonstrates a fractional frequency stability of $1\times 10^{-16}$ at short time scales and supports a laser linewidth of $<$40~mHz at 1.5 $\mathbf{\mu}$m, representing an optical quality factor of $\>4\times 10^{15}$.
 \end{abstract}

\maketitle
\section*{Introduction}
Precision optical interferometers are at the heart of the world's most accurate measuring science such as optical atomic clock work~\cite{cho10,lud08,tam09,fal11}, gravitational wave detection~\cite{har06b,wil08c,abb09}, cavity quantum electrodynamics~\cite{bir05}, quantum optomechanics~\cite{mar03c,abb09a} and precision tests of relativity~\cite{eis09}. These experiments explore the next frontiers of measurement science and open new windows for scientific discoveries. However, the ultimate limit to the sensitivity and stability of these interferometry devices is now set by the Brownian thermo-mechanical noise in the interferometer's mirrors or spacers~\cite{num04,not06,kes11}. This noise directly leads to fluctuations of the optical length of the interferometer, degrading the frequency stability of any laser stabilized to it~\cite{lud07,you99,jia11,wil08c} and impeding the achievement of quantum coherence in a macroscopic system. Consequently, different attempts, in both theory~\cite{kim08,kes11} and experiment~\cite{mil09,leg10}, have been made to lower this limit, which is currently at $\sim$$3\times 10^{-16}$ fractional instability~\cite{jia11}.
\par
Here, we present an approach capable of reducing this thermal noise limit by at least an order of magnitude. As a result of the fluctuation-dissipation theorem, the thermal noise of an interferometer made of high mechanical quality factor $Q$ material can be reduced significantly as compared to conventional glass materials. For such a high $Q$ material we have selected a silicon single crystal for both the cavity spacer and mirror substrates. Silicon has other important advantages: its superior Young's modulus effectively suppresses the sensitivity of the cavity length to environmental vibration noise. Also, aging-related frequency drifts commonly encountered in conventional glass materials are completely absent in crystalline materials, as demonstrated by cryogenic sapphire optical resonators operated at a temperature of 4.2~K~\cite{sto98,mue03b}. An all-silicon interferometer can be made insensitive to temperature fluctuations as the coefficient of thermal expansion of silicon has a zero crossing near 124~K. In this temperature range the $Q$ of silicon is orders of magnitude higher than in the conventional cavity glass materials such as Ultralow Expansion glass (ULE) or fused silica~\cite{nie06}. A similar approach has been used with relatively low-finesse optical Fabry-Perot interferometers~\cite{ric91} and has recently been proposed for gravitational wave detection~\cite{sch10e}. A remaining issue to be solved in the future is the thermal noise associated with optical coating, but its contribution can be reduced with the use of a longer cavity spacer.
\par
After presenting the detailed design of the silicon crystal cavity, we report frequency stabilization of a laser to a high finesse silicon interferometer at a wavelength of 1.5~$\mu$m where the silicon substrate is transparent. By comparison with two other lasers stabilized to state-of-the-art conventional interferometers, we demonstrate the superior performance of the silicon interferometer by achieving a laser linewidth of $<$40~mHz and short-term stability at $1\times 10^{-16}$. The corresponding optical coherence length is beyond $1\times 10^{9}$~m, on the order of the length of the proposed Laser Interferometer Space Antenna (LISA). While this performance exceeds any oscillators demonstrated to date in microwave or optical regions, we emphasize that the paradigm shift represented by the use of a crystal cavity instead of more traditional amorphous materials (such as ULE) will enable much further progress for time, frequency, and length metrology.
\begin{figure*}
	\centering
\includegraphics[width=.5\linewidth]{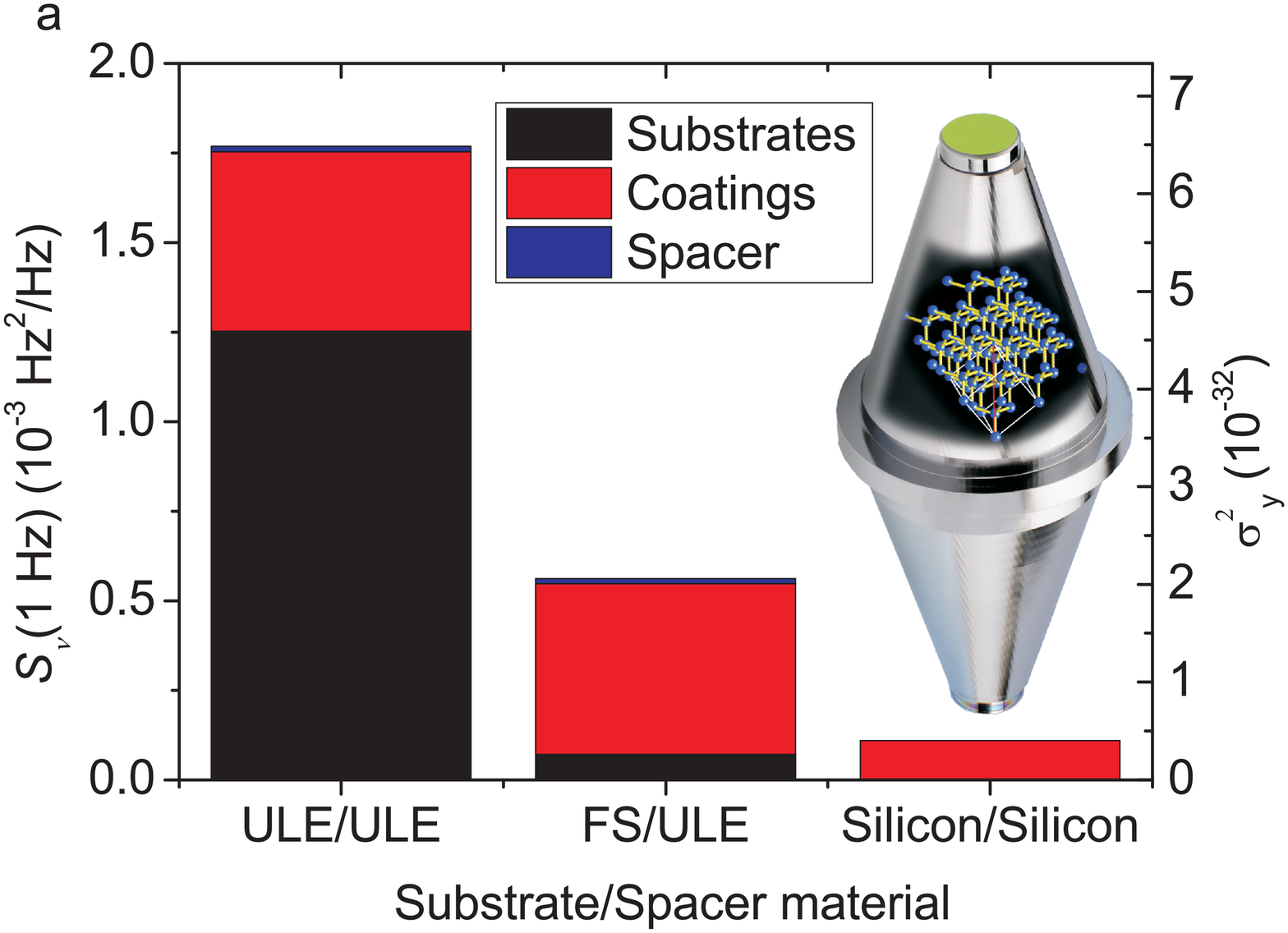}
\hfill
\includegraphics[width=.45\linewidth]{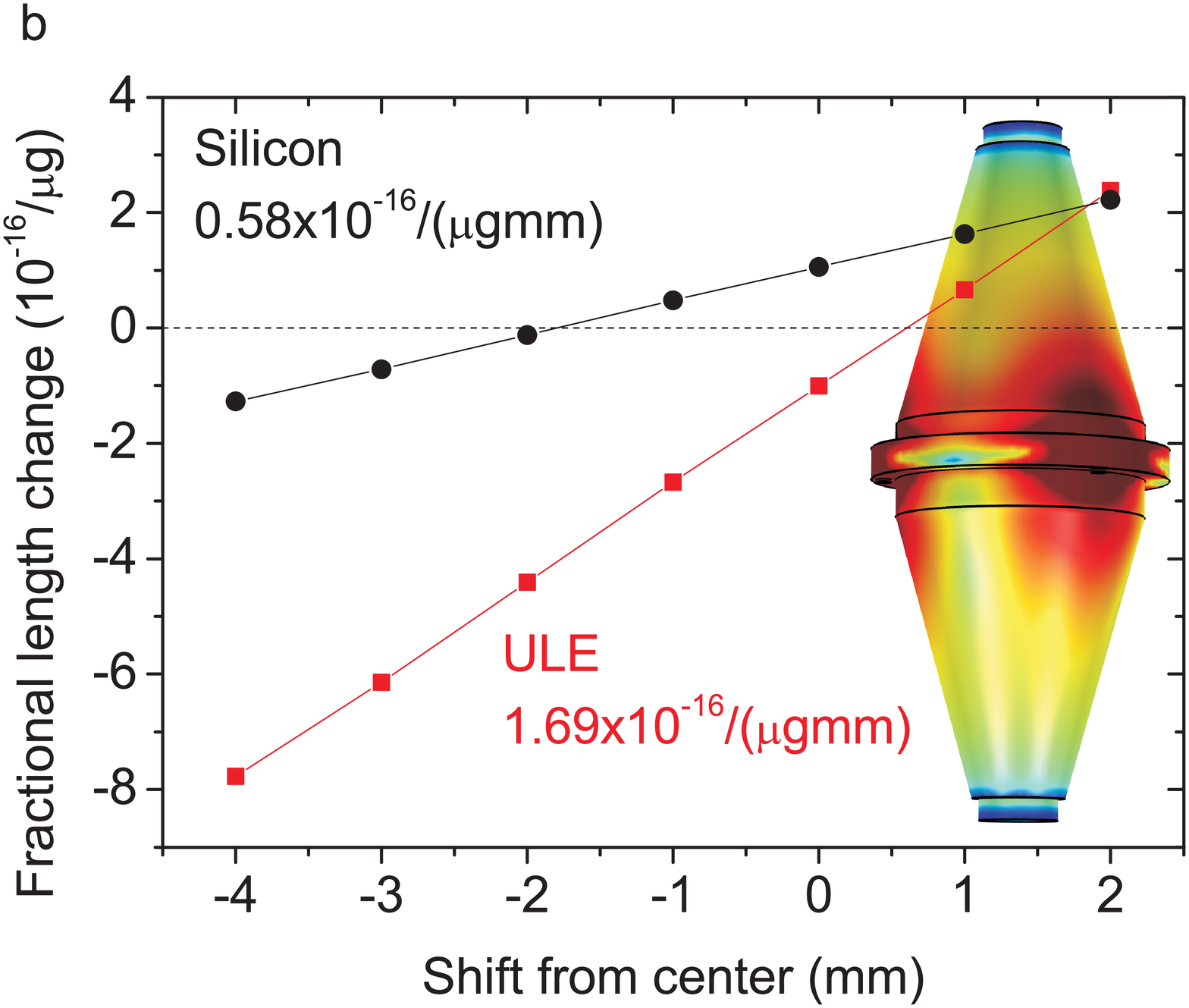} \hfill
\caption{Performance of a single-crystal silicon Fabry-Perot interferometer
(A) Estimation of frequency noise power spectral density arising from Brownian-motion thermal noise for various mirror substrate and spacer materials. The nominal dimension of the optical cavity has a spacer length of 210~mm, spacer radius 50~mm, central bore 5~mm, radii of curvature of 1 m/$\infty$ for the two mirrors. The right axis displays the corresponding Allan Variance for fractional optical frequency instability. (B) FEM simulation of the vibration sensitivity of the vertical cavity design as a function of the position of the central support ring from the symmetric plane for ULE and single-crystal silicon as cavity material. The inset shows the strain energy distribution for the silicon spacer.
}
\label{fig:performance}
\end{figure*}
\section*{Design of silicon cavity}
Mono-crystalline silicon possesses superior mechanical properties~\cite{pet82} such as high elastic modulus, well above that of optical glasses, and large thermal conductivity.
Large, high-purity and low-defect crystals are readily available at reasonable cost. Compared to glasses, the thermal conductivity (500~W/mK at 124~K~\cite{gla64}) is about two orders of magnitude higher (1.31 W/mK), which assures a homogeneous temperature and small local heating from any absorbed laser radiation. Silicon is a cubic crystal with its largest Young's modulus along the $<$111$>$ axis (E = 187.5 GPa)~\cite{wor65,bra73}, which we have chosen to be the optical axis of the cavity.
\par
Figure~\ref{fig:performance}A shows the calculated thermal noise for the silicon cavity at its zero crossing temperature of 124 K, in comparison to room temperature cavities with ULE spacers and ULE or fused silica mirror substrates. The frequency noise power spectral density of the silicon cavity-stabilized laser is reduced partially because of the lower temperature. In addition, the high mechanical quality factor of silicon ($Q > 10^7$~\cite{gui78,naw06}) directly reduces the noise contribution from the spacer and the mirror substrates.
The high Young's modulus of the mirror substrate compared to ULE and fused silica ($E \approx 67.6$~GPa) also further reduces the thermal noise of the coating~\cite{har02b}. Combining these contributions, we estimate that the thermal noise leads to a flicker floor of the fractional optical frequency stability of $\sigma_y = 7\times 10^{-17}$. In the extremely unlikely scenario where the substrate temperature is set 1~K off from the zero crossing point of the thermal expansion, the estimated thermo-optic (coating) and thermo-elastic (substrate) noise is at least two orders of magnitude below this thermal noise.

To reach this record-low level of thermal noise, we must reduce the influence of environmental vibrations on the resonator of length  $L=210$~mm. We minimized the sensitivity of the spacer length to external applied forces via finite element modeling (FEM), taking advantage of the full elasticity tensor of silicon. Designs with either a vertical~\cite{not05,lud07} or a horizontal optical axis~\cite{naz06,web07} have achieved strong suppression of vibrational sensitivity. After extensive FEM simulations, a vertical configuration similar to that of Ref.~\cite{lud07} was chosen, as it showed the smallest sensitivity to the details of the mounting (e.g., contact area and contact material). We found that $|\Delta L/L| < 10^{-16}/\mu g$ for a tolerance of the axial position of 0.5~mm, where $g=9.81$ m/s$^2$ is the gravity acceleration and $\Delta L/L$ is the fractional length change of the cavity.
This represents a factor of three improvement over a ULE cavity of the same design.

It is important to note that the anisotropic structure of silicon leads to a dependence of the sensitivity of the support points with respect to the crystal axes. We found that the $<$110$>$ directions offered the most optimized supporting configuration. The three equivalent directions of $<$110$>$ intersect with our horizontal mounting ring, giving rise to three supporting points that are symmetric by axial rotation of 120$^\circ$.  For horizontal accelerations, we find a tilt of 65~nrad/$g$, compared to 180~nrad/$g$ for ULE.
Finally, for a 10\% asymmetry in the elasticity between the mounting points we found an on-axis fractional length change of $2 \times 10 ^{-17}/\mu g$ ($5 \times 10 ^{-17}/\mu g$ for ULE). These results clearly illustrate the superior elastic properties of silicon compared to low expansion glass.

The raw material was a high resistivity silicon rod of 100~mm diameter. It was oriented by X-ray diffraction to define the optical axis to within a few arc minutes.
The crystal was machined to a tapered cylinder shape with mounting ring and the surface etched to reduce stress induced by the machining~\cite{lud07}. The silicon mirror substrates were machined from the same oriented material and, after being superpolished and coated with high-reflectivity, low-loss optical thin films, they were optically contacted to the spacer. We aligned the substrates and spacer crystal in the same crystallographic orientation as to best maintain the single crystal property of the entire cavity. We were able to obtain a finesse of 80~000 for the TEM$_\mathrm{00}$ mode of the cavity, and a much higher finesse of 240~000 for the TEM$_\mathrm{01}$ mode. Consequently, for all the results discussed in this work, the TEM$_\mathrm{01}$ mode was used for laser stabilization.

\section*{Cryostat design and performance}
\begin{figure}
	\centering
\fbox{\includegraphics[width=0.9\linewidth]{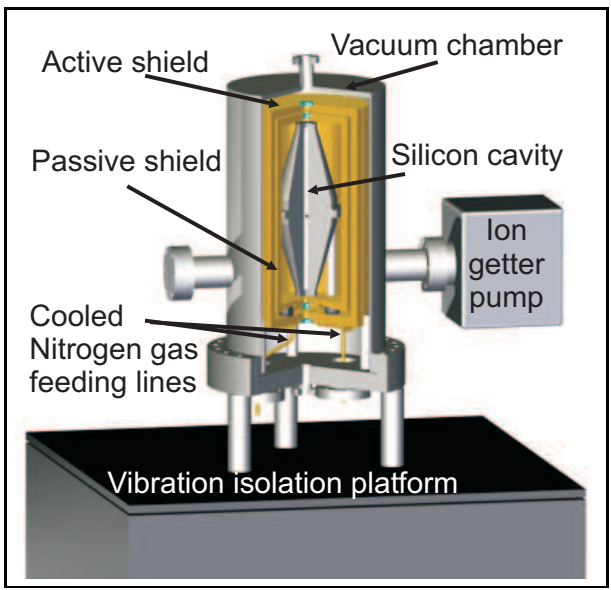}}
\caption{Schematic of the vibration-reduced, nitrogen-gas-based cryostat including the vacuum chamber and the two heat shields centered around the silicon single-crystal cavity. The thermal time constant of the system is 14 days.
}
\label{fig:cryostat}
\end{figure}
Fractional frequency stability at the level of $1\times10^{-16}$ also places a stringent requirement on cavity temperature stabilization. Thermal fluctuations are suppressed to first order by operating the cavity at a zero-crossing point of its thermal expansion, at a temperature of $\sim$124 K for silicon \cite{swe83}. A major challenge for optical cavities operated at cryogenic temperatures is the reduction of vibrations induced by the cooling mechanism. This challenge was overcome in part by a novel and simple cryostat design based on nitrogen gas as coolant. A sketch of the setup is shown in Fig.~\ref{fig:cryostat}.
The vacuum chamber is placed on an actively controlled vibration isolation table and evacuated to a residual pressure of $1 \times 10^{-8}$~mbar. The cavity is surrounded by two  massive, cylindrical, gold-plated copper-shields and supported by a tripod machined from teflon.
Cooling of the outer shield is accomplished by evaporating liquid nitrogen from a  dewar using a heating resistor. The nitrogen gas flows through superinsulated vacuum tubes to the outer heat shield and by regulating its flow the shield temperature is stabilized to maximum deviations of  $\sim$1~mK from the setpoint.
An additional thermally isolated passive heat shield suppresses radiative heat transfer from the outer shield to the silicon cavity, resulting in a thermal time constant of more than 14 days.
During continuous operation, the cryostat exhibits acceleration noise of low 0.1 $\mu g/\sqrt{\mathrm{Hz}}$ in the frequency range of 1 to 100 Hz.

 \par
The cavity length was probed by a commercial Er-doped fiber laser source.  Locking to the cavity resonance was realized via the Pound-Drever-Hall (PDH) scheme~\cite{dre83}.
Phase modulation of the laser was achieved by an all-fiber waveguide electro-optical modulator with an active suppression of the residual amplitude modulation (RAM).
Feedback was applied to a fiber-coupled acousto-optical modulator with a locking bandwidth of $\sim 200$~kHz to keep the laser tightly locked to the cavity resonance.
\par

The coefficient of thermal expansion $\alpha$ of the silicon cavity was measured in a region close to its predicted zero-crossing at 124 K by monitoring the frequency of the laser locked to the cavity as a function of the cavity temperature measured on its surface with a PT-100 sensor. The laser frequency was measured in comparison to a reference laser system stabilized at room temperature. The minimum cavity length was found at a temperature of 124.2~K with a statistical uncertainty of 5~mK, caused by the delayed response of the cavity frequency to a change of its surface temperature. The accuracy of the absolute temperature is limited to a value of 0.1~K by the calibration of the temperature sensor. We determine a zero-crossing sensitivity of $d\alpha/dT=1.71(1)\times 10^{-8}$~K$^{-2}$. These results are consistent with the values given in~\cite{swe83,gla01}.
\par
\begin{figure*}
\centering
\includegraphics[width=.95\linewidth]{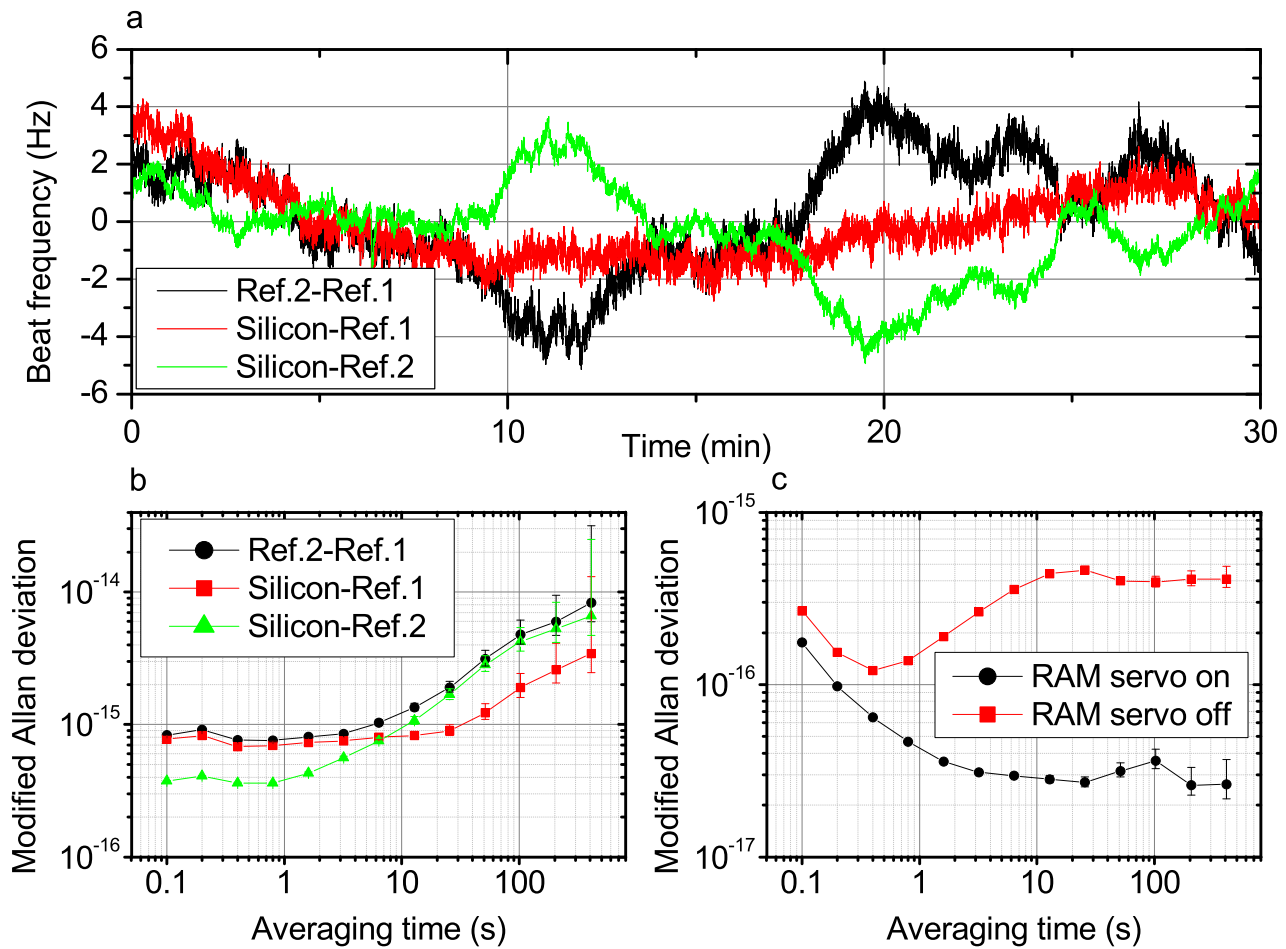}
\caption{Time record and stability of the optical beat signals between the silicon cavity laser and the two reference lasers. (A) Traces of the three beats involved in the three-cornered hat comparison after linear drift removal (Ref.2$-$Ref.1: -234.9 mHz/s, Silicon$-$Ref.1:-4.7 mHz/s, Silicon$-$Ref.2: 230.2 mHz/s. (B) Modified Allan deviation of the fractional frequency stability for the corresponding beat signals shown in panel (A). (C) Fractional frequency stability as limited by the drift of the offset of the cavity servo error signal, with and without the active RAM cancelation engaged.}
\label{fig:traces}	
\end{figure*}
To determine the cavity length sensitivity to vibrations in the horizontal and vertical directions ($k_x$, $k_y$, and $k_z$, respectively), we analyzed the frequency response of the stabilized laser system to applied excitations. We measured accelerations in all three axes while we simultaneously recorded the Si cavity-stabilized laser system with a second reference laser system. Since we observed a non-negligible coupling of the movement from the excited to the non-excited directions, we had to take into account amplitude and phase relations of all acceleration components in the full calculation~\cite{naz06}. For the horizontal directions, a fractional vibration sensitivity of $k_x=6.7(3)\times 10^{-17}/\mu g$ and $k_y=8.4(4)\times 10^{-17}/\mu g$ was determined. The vertical sensitivity is determined to be $k_z=5.5(3)\times 10^{-17}/\mu g$, hence readily enabling fractional instabilities below $10^{-16}$. The vibration sensitivity reported here is the lowest for any optical cavities designed completely based on simulations. It is only surpassed by dedicated transportable cavity designs with actively tunable degrees of freedom allowing for  experimentally minimizing the sensitivity~\cite{web11}.
\section*{Reference laser systems}
To determine the performance of the silicon cavity-stabilized laser, we utilized two of the best-performing conventional ULE-based optical cavity-stabilized lasers from JILA and PTB. Using heterodyne beats between these three independently stabilized lasers, we performed a three-cornered hat comparison~\cite{gra74} at 1.5 $\mu$m.
\par
The first reference laser (Ref.1) is designed as a compact cavity-stabilized system. A distributed feedback fiber laser is stabilized to a horizontally mounted  reference cavity made of ULE. The cavity is 10 cm long and has a finesse of 316~000.
The thermal noise of the cavity is expected to limit the Allan deviation at a flicker floor of $\sigma_y(\tau) \approx 6 \times 10^{-16}$.

\par
The second reference system (Ref.2) employs another Er-doped fiber laser frequency stabilized to a ULE cavity.  The mirror substrate material is fused silica, which lowers the magnitude of the thermally driven mirror fluctuations in comparison to a system employing ULE mirror substrates. Additionally, the cavity spacer's length of 25~cm reduces the fractional impact of these same fluctuations on the frequency stability of the cavity-stabilized laser, resulting in an expected thermal noise-limited flicker floor of $2\times  10^{-16}$. The cavity finesse is 80~000, which led to a comparatively increased sensitivity to time-dependent RAM caused by parasitic effects, making the active RAM stabilization mandatory.

To make simultaneous comparisons of the three systems, the frequency-stable light of each of the three laser systems was transferred by single-mode optical fibers to a centrally located beat detection unit. A fiber noise cancellation system~\cite{ma94,gro09} for each optical fiber path reduces fiber-noise to below $10^{-16}$ at 1~s. At the central detection unit the light of all three lasers was superimposed on 50/50 fiber-splitters. An InGaAs photodiode at the output of the last fiber-splitter detected the three beat signals.
The laser beat signals were tracked by phase-locked-loop-based filters and subsequently recorded by a dead-time free counter (K+K Messtechnik: model FXE) in phase averaging mode, which corresponds to an overlapping $\Lambda$-mode counting scheme as required for the measurement of the modified Allan deviation~\cite{rub05}. This measure for the  stability of the oscillators was chosen for obtaining the highest possible sensitivity at  averaging times greater than 100 ms where we expect to reach the thermal noise floor~\cite{daw07}.

\section*{Laser frequency stability}
To characterize the performance of all three lasers, a 24 hour long record of the beat signal was taken with a counter gate time of 100 ms.
\label{sec:adev}
Figure~\ref{fig:traces}A documents a typical 30-minute long trace of the beat signals between the three lasers under comparison after removal of the linear drift. The best long-term stability was observed for the beat between the silicon cavity-stabilized laser and Ref.1 with maximum frequency excursions of 6~Hz over the full time period. The two beat signals of Silicon$-$Ref.2 and Ref.2$-$Ref.1 show correlated fluctuations on a similar level at shorter time scales of a few minutes.
\begin{figure}
	\centering
\includegraphics[width=.95\linewidth]{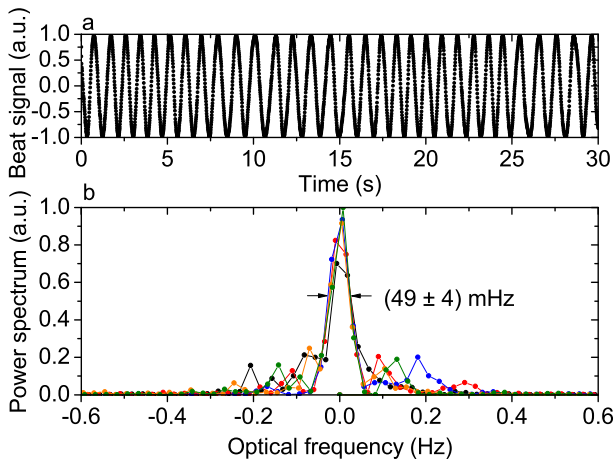}
\caption{Optical heterodyne beat between the silicon cavity system and reference laser 2. (A) Beat signal mixed down to DC and recorded with a digital oscilloscope. (B) Fast Fourier transform of the beat signal recorded with a HP 3561A FFT-Analyzer (37.5 mHz resolution bandwidth, Hanning window). Five consecutive recordings of the beat signal are displayed here, demonstrating the robustness of this record-setting linewidth. The silicon cavity stabilized laser linewidth has an upper limit of 35 mHz. }
\label{fig:linewidth}	
\end{figure}\par
For detailed analysis of the frequency stability of the three systems, modified Allan deviations~\cite{all81} were determined from the time records and are shown in Fig.~\ref{fig:traces}B. The error bars have been derived following a $\chi^2$-statistics~\cite{gre03b,ril04}.
The two laser beats involving Ref.1 are limited by flicker frequency noise to a level of $\sim 7\times 10^{-16} $, which is consistent with the thermal noise floor expected for this reference laser. The flicker frequency noise of $\sim 3.3\times 10^{-16}$ at 0.1 to 1 s averaging time for the beat between the silicon cavity-stabilized laser and Ref.2 can be attributed to a large extent to the thermal noise limit of Ref.2. The effect of active suppression of the influence of RAM to the baseline of the servo error signal for the silicon cavity system is demonstrated in Fig.~\ref{fig:traces}C.

\par
The narrowest optical linewidth was observed for the beat between the silicon laser and Ref.2.
This signal was analyzed by a digital oscilloscope (Fig.~\ref{fig:linewidth}A) and a fast-fourier-transform analyzer (Fig.~\ref{fig:linewidth}B). We obtain a typical full width at half maximum of $49(4)$~mHz which is the narrowest spectrum of a beat between two lasers observed. Considering that the silicon cavity-stabilized laser has the best stability among all three systems, it is a reasonable estimate that its linewidth is below 35 mHz. The optical coherence time is 10~s.

\begin{figure}
\centering
\includegraphics[width=.9\linewidth]{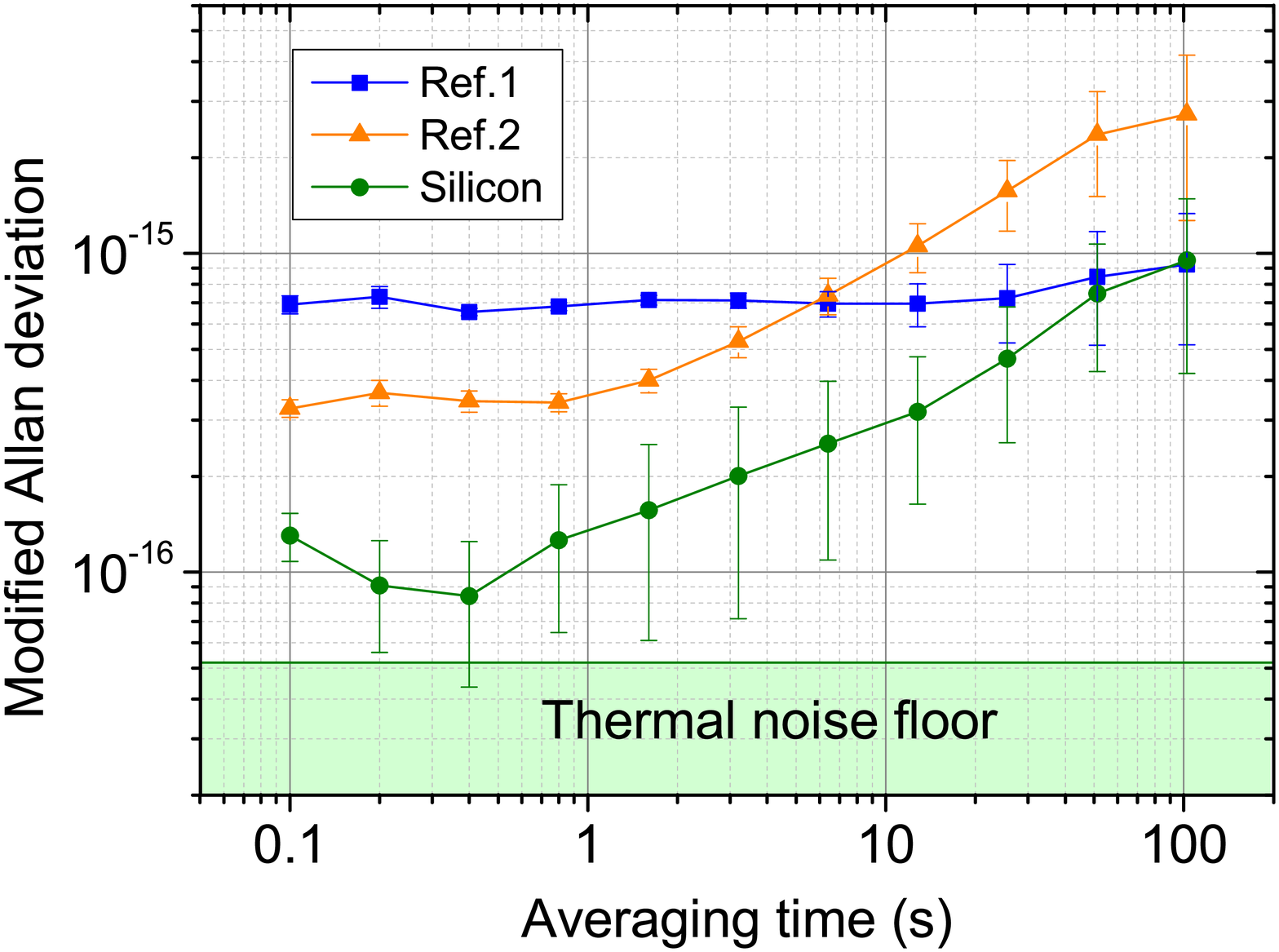}
\caption{Modified Allan deviation of each cavity-stabilized laser derived from a three-cornered-hat analysis of the data in Fig.~\ref{fig:traces}. The data shows the average stability of 130 data-sets with a duration of 10 minutes each. The error bars represent the standard deviation of the data averaged for each point. The predicated thermal noise floor of $\sim 5\times10^{-17}$ for the silicon cavity system is indicated by the shaded area.}
\label{fig:madevsingle}	
\end{figure}	
\par
To evaluate the performance of each single laser, rigorous three-cornered hat analysis including correlations between the three oscillators was carried out following~\cite{pre93}. To verify the reproducibility of  the analysis the full data-set was split into 130 data-sets with a duration of 10 minutes each.
For those data-sets three-cornered hat analysis was performed including the effect of possible correlations between the laser systems~\cite{pre93}. The average stability of each of the three lasers is shown in Fig.~\ref{fig:madevsingle}. The error bars for the single oscillators represent the standard deviation of the mean values.
\par
At short averaging times the two reference lasers are limited to within a factor of two of their thermal noise floors of $\sim6\times 10^{-16}$ and $\sim2\times 10^{-16}$, respectively. The predicted thermal noise for the silicon cavity-stabilized laser of $\sim 5\times10^{-17}$ is indicated in the graph. Note that instabilities arising from flicker frequency noise in a modified Allan plot are reduced by $\sim 18$\% in comparison to the commonly used standard Allan deviation as pointed out in~\cite{daw07}.
The stability of the laser locked to the silicon cavity clearly surpasses the performance of the two ULE-cavity-based reference lasers and consequently the silicon data contains the largest statistical uncertainty. The silicon cavity system reaches residual fractional frequency instability of $\sim1\times10^{-16}$ for averaging times of 0.1-1~s and remains at the low $10^{-16}$ level for averaging times up to 10~s, which is a time scale relevant for long-term stabilization to the strontium atomic frequency standards~\cite{swa11} available at both JILA and PTB. The instability of the silicon cavity system at short time scales (below 0.5~s) is limited by the measured vibration noise floor and RAM. This level of performance is the best fractional frequency stability of any oscillator, optical or microwave, to date. Such a stable cavity will have direct impact on optical atomic clocks and will also have important implications for quantum metrology and precision measurement science in general.

\section*{Outlook}
Our first investigations of silicon as a material for an optical reference cavity with ultra-low thermal noise has already demonstrated performances that surpass any previously reported optical cavities. At this early stage, a number of technical challenges are yet to be mastered to reach the predicted thermal noise floor of $\sim5\times 10^{-17}$ and to further lower this limit. RAM errors in cavity servos can be suppressed to the required level by active feedback. However, at the short time scales the vibration noise of the cryostat should be further reduced. The mechanism that limits the stability to $\sim1\times 10^{-16}$ for averaging times of 1--100~s needs further investigations based on a direct comparison with a second silicon cavity-stabilized laser  we are currently implementing.
\par
Material creep is not expected from a single-crystal cavityaging
 and therefore the cavity will also serve as an excellent long-term stable frequency reference, possibly being competitive with the stability of a hydrogen maser at timescales of 1000~s and beyond. Long-term measurements against a frequency comb locked to the primary cesium standard~\cite{lip09} will reveal if this is indeed the case. Our preliminary experiments are already recording a long-term optical frequency drift of only a few mHz/s, but longer data sets will be required for detailed analysis. In combination with an optical frequency comb, which allows the stability of the silicon system to be transferred to any wavelengths within the comb spectral coverage~\cite{tel02b,sch08a}, the infrared ultra-stable laser will provide a powerful tool for high-resolution spectroscopy and optical clocks. With clock instabilities of  $1\times 10^{-16}/\sqrt{\tau}$ that will be possible with such a laser as local oscillator,
 observation times needed to resolve frequency shifts at a level of $10^{-17}$ would be at the 100~s level~\cite{swa11}.
\par
Finally, we note that the thermal noise limit of an optical cavity based on single-crystal silicon is currently set by the optical coating. We plan to investigate different approaches to push down this limit. Possible candidates for the next generation optical coatings are microstructured gratings~\cite{bru10} or mirrors based on III/V materials such as gallium-arsenide~\cite{col08}. With these materials an instability approaching $~10^{-17}$ and below could be within reach.

\section*{Acknowledgement} 
This silicon cavity work is supported and developed jointly by the Centre for Quantum  Engineering and Space-Time Research (QUEST), Physikalisch-Technische Bundesanstalt (PTB), and the JILA Physics Frontier Center (NSF) and the National Institute of Standards and Technology (NIST).  We thank R. Lalezari and Y. Lin of ATF for the coating of the silicon mirrors and the initial finesse measurements. We also thank M. Notcutt and R. Fox for the technical assistance for the construction of the second reference cavity. U. Kuetgens and D. Schulze are acknowledged for x-ray orientation of the spacer and the mirrors. G. Grosche is acknowledged for technical assistance on fiber noise cancellation.  J. Ye thanks the Alexander von Humboldt Foundation for financial support.

 \bibliographystyle{prsty}
 \bibliography{siscience}

\end{document}